\documentstyle[12pt]{article}
\begin{document}
\title{Quantum Spaces: \\
       Notes and comments \\on a lecture by S. L. Woronowicz
\thanks{Supported by KBN under grant 2 1047 91 01 } }
\author{R. J. Budzy\'nski\thanks{email: budzynsk@fuw.edu.pl}
\\ Department of Physics, Warsaw University
      \and
        W. Kondracki\thanks{email: witekkon@impan.gov.pl}
\\ Institute of Mathematics, Polish Academy of Sciences}
\date{IM PAN 516 \\ hep-th/9401018 \\December 1993}
\maketitle
\begin{abstract}
These notes are based on a lecture given by S. L. Woronowicz
at the Institute of Mathematics, Polish Academy of Sciences.
\end{abstract}
\section*{Introduction}
The concept of quantum spaces has been under study for some time now, and
recently it has been attracting a growing amount of attention. This is
in part due to the interest in quantum groups, which are considered a very
promising concept by many, both physicists and mathematicians. It seems
that quantum groups appear quite naturally in low-dimensional quantum field
theory and statistical mechanics, they have also found important applications
in the theory of knots. Other potential hoped-for applications include a
possible role of quantum spaces in a future quantized theory of gravity.

The theory of quantum spaces, and quantum groups in particular, is at present
in a phase of rapid growth. One of the consequences of this is that the
language used in the field still remains somewhat ambiguous, even
concerning the definition of basic concepts. Here, the category of locally
compact quantum spaces will be understood as the dual to the category of
$C^\star$ algebras \cite{Dix}.
The main objective of the present notes is to motivate
and describe this basic notion.

\section*{Compact quantum spaces}
Let us begin by considering a compact Hausdorff space $X$, and the set
$C(X)$ of continuous, complex valued functions on $X$. $C(X)$ is naturally
endowed with the structure of a commutative algebra with unit over the complex
number field, equipped moreover with the anti-linear involution $\star$ given
by
\[ (f^\star)(p) = \overline{f(p)}\]
 and the norm
\[ \| f\| = \sup_{p \in X}|f(p)|. \]
This norm can be seen to obey the condition
\[ \|f^\star f\| = \|f\|^2. \]
It is
also known that the space $C(X)$ is complete with respect to this norm.

Algebras (not necessarily commutative) with the above properties are known
as $C^\star$ algebras. We see therefore that every compact Hausdorff space $X$
is in a natural way associated with a commutative $C^\star$ algebra with unit,
namely $C(X)$. What's more, every continuous mapping between compact
Hausdorff spaces, $T : X \rightarrow Y$, determines a $C^\star$ homomorphism:
$T^\star : C(Y) \rightarrow C(X)$, given by
\[ (T^\star f)(p) = f(T(p)). \]

In particular, when $X$ consists of a single point, $C(X) = {\bf C}$ and
any mapping of $X$ into $Y$ determines a linear and multiplicative
functional on $C(Y)$. In brief: points of $Y$ correspond to linear
multiplicative functionals on $C(Y)$. The celebrated Gelfand--Naimark
theorem states that this correspondence is one to one. This leads to
an important conclusion: Any commutative $C^\star$ algebra with unit is the
algebra of continuous functions on a compact topological space. This space
may be identified with the set of linear multiplicative functionals
(also known as characters) on the algebra.

Stating it in the language of category theory: there exists a contravariant
isomorphism between the category of compact topological spaces and that
of commutative $C^\star$ algebras with unit. Consequently, it is natural to
make the following generalization: A compact {\em quantum} space corresponds,
by an extension of this isomorphism, to a {\em noncommutative} $C^\star$
algebra with unit. It must be stressed that it is not necessary (nor does
it seem useful) to consider the quantum space itself, as a set.

It may be therefore said that the theory of quantum spaces is no more
than the theory of $C^\star$ algebras. Such a point of view, however, misses
much of the point. A similar example is the theory of probability:
one could just as well say that it is no more than measure theory. The
fact is that although both probability and measure theory may be reduced
to a common denominator, they differ in the kind of questions they ask
and in their applications.

Returning to the main subject, one of the basic concepts of topology is
that of the cartesian product of topological spaces. In the language
of $C^\star$ algebras, the corresponding concept is that of the tensor
product of algebras. This, however, requires a definition of the tensor
product that would operate within the category of $C^\star$ algebras. The
problem boils down to properly defining the norm and completion of the
algebraic tensor product. To show how this is done, we quote the following
theorem due to Gelfand, Naimark and Segal:
\begin{quote}
Every separable $C^\star$ algebra
has a faithful continuous representation in $B(H)$, the algebra of bounded
linear operators on a Hilbert space $H$. Moreover, this representation
is isometric (norm-preserving).
\end{quote}
It follows that the image of such a
representation is a closed subalgebra of $B(H)$.

We say that a closed subalgebra $A \subset B(H)$ is nondegenerate iff
for every nonzero $x \in H$, $ax \neq 0$ for some $a \in A$. The family
of separable, nondegenerate subalgebras of $B(H)$ will be denoted as
$C^\star(H)$.

We are now ready to explain the construction of an adequate tensor product
of $C^\star$ algebras: Consider the $C^\star$ algebras $A$ and $A'$.
In virtue of
the Theorem quoted above, $A$ and $A'$ may be identified with certain
elements of $C^\star(H)$. The algebraic tensor product is contained in
$B(H) \otimes B(H) \subset B(H \otimes H)$. Completing the image of
$A \otimes A'$ with respect to the norm in $B(H \otimes H)$, we obtain
a separable $C^\star$ algebra, which we identify as the tensor product of
$A$ with $A'$. It turns out that the resulting algebra does not depend
on the choice of (faithful) representations of $A$ and $A'$.

An important property of the cartesian product is the existence of natural
projections onto the factors:
\[ X \times Y \ni (x,y) \longmapsto x \in X ,\]
\[ X \times Y \ni (x,y) \longmapsto y \in Y .\]
Moreover, in the case of the cartesian product of a space with itself
there exists a natural {\em diagonal} mapping:
\[ X \ni x \longmapsto (x,x) \in X \times X .\]
In the language of quantum spaces, the projections correspond to the natural
injective homomorphisms:
\[ A \ni a \longmapsto a \otimes I \in A \otimes A' ,\]
\[ A' \ni a' \longmapsto I \otimes a' \in A \otimes A' .\]
The diagonal mapping, however, does not have a quantum equivalent: the
mapping
\[ A \otimes A \ni \sum a_i \otimes b_i \longmapsto \sum a_i b_i \]
is not an algebra homomorphism in general.

\section*{Examples}
Many examples of quantum spaces may be presented as algebras with
a finite number of generators and relations. Consider such a finite set
of generators $(I, a_i)$ and relations $R_k$. We form a $\star$-algebra with
unit, $A$, as the quotient of the free algebra in the generators $(I, a_i)$
by the two-sided ideal generated by the relations $R_k$. Next, we consider
the set $\{\pi\}$ of $\star$-representations of $A$ in $B(H)$.
If for every $x \in A$,
$\sup_{\pi} \|\pi(x)\|$ exists, it determines a semi-norm in $A$. By
quotienting $A$ additionally by the ideal generated by elements of vanishing
semi-norm and completing, we obtain the so-called universal $C^\star$ algebra
generated by $(I, a_i)$ with relations $R_k$.

As an example, we now present the {\em quantum disk}\cite{Klim}.
Let $\mu$ be a real
number, $0< \mu <1$, and $z$ a generator subject to the relation
\[ zz^{\star} - z^{\star}z = \mu (I - z^{\star}z)(I - zz^{\star}) .\]
It turns out that such an algebra admits only two types of inequivalent
representations:
\begin{itemize}
\item $\pi(z) \in {\bf C},\; |z| = 1$,
\item $\pi(z)f_n = (\frac{n\mu}{1 + n\mu})^{1/2} f_{n-1}$,
\end{itemize}
where $(f_n)_{n=0}^{\infty}\:$ form an orthonormal basis
in the Hilbert space $H$.
The algebra $A_{\mu}$ described here is well known as the
{\em Toeplitz algebra}.

Observe that the mapping
\[ z \longmapsto e^{i\phi} \in C(S^1) \]
extends to
a homomorphism of $A_{\mu}$ onto $C(S^1)$. In this sense, the usual
circle is a {\em subset} of the quantum disk.

The following nontrivial fact can be shown: the group $SU(1,1)$ of two by two
complex matrices
\[ \left( \begin{array}{cc}
\alpha & \beta \\
\gamma & \delta
\end{array} \right)
\]
such that $ \beta = \overline{\gamma}, \delta = \overline{\alpha}$, and
$|\alpha|^2 - |\beta|^2 = 1$, acts on $A_{\mu}$ as an automorphism group:
\[ \zeta = (\alpha z + \beta)(\gamma z + \delta)^{-1} \in A_{\mu} \]
satisfies
the same relation as $z$; this justifies the term quantum disk, being in
precise analogy with the action of $SU(1,1)$ on the unit disk in $\bf C$,
as its group of biholomorphisms.

What's more, it is possible to contemplate
a quantum extension of a well-known construction of closed Riemann
surfaces of genus $> 1$. Any such surface $\Sigma$ can be presented as
the quotient of the disk by the action of a discrete subgroup of $SU(1,1)$,
isomorphic to $\pi_1(\Sigma)$. The quantum version of this would be
to consider the subalgebra
\[ A_{\Sigma} =
\{x \in A_{\mu} : \varphi_g(x) = x {\rm \; for\; all\;}
g \in \pi_1{\Sigma}\},\]
where $\varphi$ denotes the $SU(1,1)$ action on $A_{\mu}$.

To make this precise, it would however be necessary to take into account
that one should be quotienting the {\em open} disk, while $A_{\mu}$ is
a quantum version of the {\em closed} disk with boundary. Before this can
be done, one needs a notion of noncompact quantum space.

Another example which we shall present is the quantum sphere
$S^3_q$\cite{SLW1}.
Let $\alpha$ and $\gamma$ be the generators, $q$ a real parameter
$(0 < q \leq 1)$, and take the following set of relations:
\[ \alpha \gamma = q \gamma \alpha, \]
\[ \alpha \gamma^{\star} = q \gamma^{\star} \alpha, \]
\[ \gamma \gamma^{\star} = \gamma^{\star} \gamma, \]
\[ \alpha^{\star} \alpha + \gamma^{\star} \gamma = I ,\]
\[ \alpha \alpha^{\star} + q^2 \gamma^{\star} \gamma = I. \]
When $q$ is set to $1$ we obtain a commutative algebra; by inspecting the
last two relations we see that, after suitable completion, it is the algebra
$C(S^3)$. In the general case, the universal $C^{\star}$ algebra
obtained from this set of generators and relations by the previously described
procedure will be called {\em the algebra of functions on the quantum sphere},
$C(S^3_q)$.

Just as is the case for the classical sphere (i.e. at $q=1$), also for
$q \neq 1$ we can find an action of $S^1$ on $C(S^3_q)$ by $C^{\star}$-algebra
automorphisms. This action is determined by the formulas
\[ S^1 \ni e^{it} \longmapsto \sigma_t : C(S^3_q) \rightarrow C(S^3_q),\]
\[ \sigma_t (\alpha) = e^{it}\alpha,\; \sigma_t (\gamma) = e^{-it}\gamma .\]
In the algebra $C(S^3_q)$ we may distinguish the set of elements left
invariant by the action $\sigma$. It forms a $C^{\star}$ algebra with unit,
and therefore corresponds to a certain compact quantum space. In the case
$q = 1$ this algebra is $C(S^2)$, and the construction we have outlined
is known as the Hopf fibration: $S^2 \sim S^3/S^1$. What we have obtained
is therefore a quantum generalization of the Hopf fibration\cite{Pod1}.

It must be mentioned that choosing the particular form of
``quantum deformation''
of $S^3$ displayed above has a deep justification. It is known that
the classical sphere $S^3$ is itself endowed with the structure of a
(topological) group;
this is the group $SU(2)$ of unitary two by two complex matrices of
determinant $1$. We have claimed that any topological notion referring
to compact spaces can be translated into the language of commutative
$C^{\star}$ algebras. In particular, a group multiplication is simply a
continuous mapping from $G \times G$ to $G$, subject to certain additional
conditions (namely associativity, existence of a unit and of inverse elements).
Speaking in terms of $C(G)$, what corresponds to group multiplication on $G$
is a (unital $C^{\star}$) homomorphism
\[ \Delta : C(G) \rightarrow C(G)\otimes C(G), \]
called the coproduct. Explicitly:
\[ C(G) \ni f \longmapsto \Delta(f) \in C(G)\otimes C(G)\sim C(G \times G)\]
is given by
\[ (\Delta (f))(g_1,g_2) = f(g_1g_2). \]
Of course, the coproduct obeys certain properties derived from the axioms
of group multiplication. We may take these as the basic axioms, and,
in the spirit of the theory of quantum spaces,
drop the requirement that $C(G)$ be commutative. This
leads to the theory of quantum groups\cite{SLW2},
which are a very important special
class of quantum spaces.

In particular, the algebra $C(S^3_q)$ described above provides an example
of a quantum group, denoted $SU_q(2)$. The coproduct is determined by
the formulas
\[ \Delta (\alpha) = \alpha \otimes \alpha - q \gamma^{\star} \otimes \gamma,
\]
\[ \Delta (\gamma) = \gamma \otimes \alpha + \alpha^{\star} \otimes \gamma .\]
It may be verified that $\Delta$ fulfills the axioms dual to those of
group multiplication.

The quotient construction of the quantum sphere $S^2_q$ sketched above
has therefore also provided us with an example of a {\em quantum homogeneous
space} for the quantum group $SU_q(2)$\cite{Pod2}.

\section*{Remarks on noncompact quantum spaces}
Up to now, we have been considering the category of {\em compact} quantum
spaces, motivating its definition by the classical version of the
Gelfand--Naimark theorem. To proceed to the case of locally compact, but
noncompact spaces, we quote a generalization of this theorem:
\begin{quote}
Every commutative $C^{\star}$ algebra may be identified with an algebra
of continuous functions on a locally compact topological space. If the
algebra has a unit element, this space is moreover compact. In the
contrary case, we are dealing with the algebra of continuous functions
on a noncompact space, subject to the condition of vanishing at infinity.
\end{quote}
Denote the algebra of continuous functions vanishing at infinity on
a noncompact space $X$ by $C_{\infty}(X)$. It turns out that in the
noncompact case there is a problem with properly defining the morphisms
of function algebras. One may easily find a continuous
mapping $F: X \rightarrow Y$ between two topological spaces, such that the
corresponding dual mapping $F^{\star}$ is {\em not} a homomorphism between
$C_{\infty}(Y)$ and $C_{\infty}(X)$; that is, $F^{\star}$ might not preserve
the condition of vanishing at infinity. $F^{\star}$ may indeed be defined, as
a homomorphism from $C_b(Y)$ to $C_b(X)$, where $C_b$ denotes the algebras
of all continuous {\em bounded} functions.
Trying to describe a noncompact topological space $X$ by its algebra of
all bounded continuous functions $C_b(X)$ is not, however, the right idea:
the latter is a $C^{\star}$ algebra
with unit, and therefore (by Gelfand--Naimark) corresponds uniquely to
a certain {\em compact} topological space (namely, the \v{C}ech--Stone
compactification of X).

To describe a solution to the problem of uniquely associating with any
locally compact topological space a certain function algebra, {\em and}
an algebra morphism with any continuous mapping, we point out the following
fact: knowing $C_{\infty}(X)$ allows one to determine $C_b(X)$. Indeed,
\[ f \in C_b(X) \Rightarrow \; {\rm for\; every\;} g \in C_{\infty}(X),\;
fg \in C_{\infty}(X). \]
In fact, $C_b(X)$ is the largest algebra which contains $C_{\infty}(X)$
as a ``sufficiently large'' ideal. By this we mean the following property:
for any $a \in C_b(X)$ which is a strictly positive function, $aC_{\infty}(X)$
is dense in $C_{\infty}(X)$. We therefore define the morphisms in the
category of commutative $C^{\star}$ algebras in the following way:
\[ F^{\star} \in {\rm Mor}(C_{\infty}(Y),C_{\infty}(X))\; {\rm iff}\;
F^{\star}:C_{\infty}(Y)\rightarrow
C_b(X)\; {\rm is\; a\; homomorphism},\]
and $F^{\star}(C_{\infty}(Y))C_{\infty}(X)$ is dense
in $C_{\infty}(X)$.

We are now ready to describe the ``quantum extension'' of this construction.
Let $A$ be a $C^{\star}$ algebra, not necessarily with unit. By the second
Gelfand--Naimark theorem, there exists a faithful representation of
$A$ in $B(H)$, i.e. $A \in C^{\star}(H)$. We treat $A$ as playing the role of
$C_{\infty}(X)$, while for that of $C_b(X)$ we take the so-called
algebra of multipliers of $A$, $M(A)$:
\[M(A) = \{a \in B(H)\; :\; aA, Aa \subset A \}. \]
For a pair of $C^{\star}$ algebras $A, B$ we will define Mor$(A, B)$ as the
class of $C^{\star}$-homomorphisms $\varphi : A \rightarrow M(B)$ such that
$\varphi(A)B$ is dense in $B$. It turns out that any such homomorphism
extends in a canonical way to a homomorphism from $M(A)$ to $M(B)$, and
the composition of morphisms can be defined.

It therefore remains to interpret, within the language of algebras,
the notion of a continuous, not necessarily bounded function. What
motivates the suitable quantum extension in this case is the following
property, which holds for continuous functions: if $f$ is continuous
on a locally compact space $X$, then $f(I + f^{\star}f)^{-1/2}$ is continuous
and bounded on $X$. This leads us to introduce the following definition:
given an algebra $A \in C^{\star}(H)$, an operator $T$ (in general unbounded)
on $H$ will be called affiliated with $A$ if
\[ T(I + T^{\star}T)^{-1/2} \in M(A), \]
and $(I +T^{\star}T)^{-1/2}A$ is a dense set in $A$. The set of affiliated
operators is to be considered as the quantum version of the set of
continuous, not necessarily bounded functions on a locally compact space
\cite{SLW3}.
The operators affiliated with an algebra do not, however, in general
themselves form an algebra.

\section*{Concluding remarks}
What we have described in the present notes is of course only the point
of departure for the theory of quantum spaces; at present, this theory is
at a rather early stage, and much remains to be developed. It does,
however, seem possible to translate into the language of $C^{\star}$ algebras
most of the notions and constructions of classical topology; for instance,
some amusing examples of surgery on quantum spaces have been displayed.
Another promising direction is the study of quantum spaces equipped with
additional structures. Here the most studied example are compact matrix
quantum groups, though interesting examples of noncompact
quantum groups have also
been constructed\cite{PW,SLW4}.
Perhaps the most interesting problem is that of finding
a satisfactory formulation of the notion of differential structure on
a quantum
space\cite{Con}(some examples can be found in \cite{SLW1,Pod3,SLW5}).\\

The authors wish to thank Professor S. L. Woronowicz for providing
an outline of his lecture.

\end{document}